\documentclass[twocolumn,letterpaper]{IEEEAerospaceCLS} 
\usepackage{amsmath,amssymb,amsfonts}
\usepackage{amscd}
\usepackage{url}
\usepackage{graphicx}
\newcommand{\ignore}[1]{}  
\newtheorem{defn}{Definition}

\begin{document}

\title{ 
Stress Propagation in Human-Robot Teams Based  on Computational Logic Model	\thanks{}
	}
	
\author{%
 Peter Shmerko and Svetlana Yanushkevich\\
ESE Dept, Schulich School of Engineering\\
University of Calgary\\
Canada\\
\and
Yumi Iwashita and Adrian Stoica\\
Jet Propulsion Laboratory\\
Pasadena, CA, USA\\
}

\maketitle

\thispagestyle{plain}
\pagestyle{plain}

\maketitle

\thispagestyle{plain}
\pagestyle{plain}

\begin{abstract}
Mission teams are exposed to the emotional toll of life and death decisions. These are small groups of specially trained people supported by intelligent machines for dealing with stressful environments and scenarios. We developed a composite model for stress monitoring in such teams of human and autonomous machines. This modelling aims to identify the conditions that may contribute to mission failure. The proposed model is composed of three parts: 1) a computational logic part that statically describes the stress states of teammates; 2) a decision part that manifests the mission status at any time; 3) a stress propagation part based on standard Susceptible-Infected-Susceptible (SIS) paradigm. In contrast to the approaches such as agent-based, random-walk and game models, the proposed model combines various mechanisms to satisfy the conditions of stress propagation in small groups. Our core apporach involves data structures such as decision tables and decision diagrams. These tools are adaptable to human-machine teaming as well. 
\end{abstract}


\section{Introduction}




Stress contagion refers to the transfer of a stressed state from one individual to another through observation or interaction. Emotional-contagion is a process that involves expressing and feeling the emotions that are similar to those of others \cite{[Hill-2010]}. Team leaders may benefit from being aware of the emotional dynamics of team members.
In their missions, first-responders and tactical operators such as firefighters, special forces, tactical police officers (i.e., SWAT) experience challenges when teaming in harsh environments (e.g., high levels of radiation, high explosive risk, extreme temperatures or pressures, lack of oxygen, low visibility because of smoke or darkness \cite{[Alon-2021],[Hagemann-2022],[Harris-2017],[Kelley-2019]}). These and other factors impact the operators' performance. For example, thermal stress degrades the cognitive functioning including response time, vigilance, and sustained attention \cite{[Hancock-2020],[Lopez-Sanchez-2018]}.

Teaming and the related stress dynamics is further complicated in the case of mixed human and autonomous machine (robots) operations. Human–robot teaming is needed for extending the operational capabilities of first-responders \cite{[Mandrake-2022]}, firefighters \cite{[Roldan-2021]}, and in the military or aerospace context, given that robots can endure harsh environments and prolonged periods of operation \cite{[Lin-2022]}. Autonomous machines are often used in aeronautics (e.g., NASA science-driven missions \cite{[Mandrake-2022]}).  
During emergency missions, humans and autonomous machines must operate under high pressure and rapid response times. In these  human-robot relations, robots should be viewed as team members, rather than tools.

In this paper, a team is defined as two or more individuals who socially interact, possess one or more common goals, and have different roles and responsibilities to complete a mission. 
In a small team, the stress state of each combatant and their teaming functions poses critical risk to the overall goals of the mission \cite{[Dietz-2017],[Gamble-2018],[Hagemann-2022]}. 
From a taxonomical point of view, the teaming process includes coordination, monitoring, planning, and affect management. If not mitigated by adaptive coping, stress reduces teammate support through attentional narrowing and inattention to others. The need in such mitigation is the motivation of this paper.

This paper is organized as follows. In Section \ref{sec:Terminology}, we  provide the terminology. Related works are reviewed in Section \ref{sec:Related-works}. The problem is formulated in Section \ref{sec:Framework-approach-problem-formulation}, and the proposed methodology and model are explained in Sections \ref{sec:formalization} and \ref{sec:model}. In Section \ref{sec:guidelines}, the computational guidelines are introduced. We discuss the results in Section \ref{sec:Discussion} and provide conclusion in Section \ref{sec:conclusions}.

\section{Terminology}\label{sec:Terminology}

\begin{itemize}
\item []\hspace{-2mm}\emph{Combat scenario} -- a set of possible scenes to achieve the mission goal. 
\item []\hspace{-2mm}\emph{Teammate} -- a human or autonomous machine. 
\item []\hspace{-2mm}\emph{Stress response function} -- a response from an autonomous machine to a stressed human teammate.
\item []\hspace{-2mm}\emph{Decision table} -- a representation of a team mission scenario using a set of scenes and the corresponding scene-decisions.
\item []\hspace{-2mm}\emph{Decision tree} -- a graphical representation of a decision table.
\item []\hspace{-2mm}\emph{Decision diagram} -- an optimized decision tree.
\item []\hspace{-2mm}\emph{Factor graph} -- a visual tool for the decomposition of a high-dimensional joint posterior distribution.
\end{itemize}

\section{Related works}\label{sec:Related-works}

In this section, we review the propagation paradigm that was studied, particularly in social sciences: epidemiology and genomic theory. We consider the stress propagation to be a particular case of this general propagation mechanism.

\subsection{Key trend -- composite models}

Modelling and simulation is a common way of understanding
and predicting the behaviour and performance of  groups of individuals \cite{[Dietz-2017]}. Various approaches and methodologies for modelling the crowd behaviour, in particular, were considered in \cite{[Li-2019],[Urizar-2016],[Veld-2015]}. Integration of an agent-based model into an epidemic model provides useful insights \cite{[Fast-2015]}.
Such mixed models  allow modellers to describe individual behavioural and
psychological characteristics, as well as  propagation between agents. These models are theoretically
able to describe a variety of different factors related to
human behaviour, such as gender, age, cultural variation,
emotional characteristics, etc.

The previously reported studies on stress propagation did not offer any fully developed models. There are, however, few well-identified trends in various fields  that can be used to form a composite approach, i.e., a combination of different modelling tools. Those trends and approaches include rumour propagation \cite{[Agarwal-2022]},  network propagation in gene modelling \cite{[Karlebach-2013],[Leifeld-2018],[Zobolas-2022]}, and the study of the impact of social media during pandemics \cite{[Fast-2015]}.
In \cite{[Munoz-2021]}, stress is considered to be a 3-factor composite model:  1) an event
stress factor (high volume of work), 2) a time pressure factor (limited time resource needed to complete the pending tasks), and 3) an effective fatigue factor (accumulated tiredness).

  

\subsection{Specific features of stress propagation}

	Despite similarities to other behaviour propagation processes, stress propagation in  a first-responder team is characterized as follows:
		\begin{itemize}
		\item [$-$]an affective state of a team member or risk of such state is assessed  (detected) using  personal wearable devices;
				\item [$-$]the affective state propagation impacts only human team mates; the autonomous machines  can  detect and report a state of a human;
\item [$-$]the behaviour of each team member and their ``cooperation'' determines the success of a mission; and
\item [$-$]the stress assessment is based on the personalized data gathered from  previous missions or training.
	\end{itemize}
	

\subsection{Robots as tools vs. team mates}

The use of  robots as tools is a common practice in operation for first-responders and tactical operators,    (e.g., drones and  drone swarms in fire-fighting \cite{[Alon-2021],[Roldan-2021]}). Once the level of machine intelligence becomes more sophisticated, the role of the machine is  shifted from the automated tool category  to the teammate  category   \cite{[Seeber-2020],[Lawless-2022],[Crandall-2018]}. This paradigm shift drives the need for the new interaction strategies between machines and humans. 
Autonomy differs from conventional automated systems in the ability of the system to
monitor its state and make decisions on its own. The key research question in many studies is  ``How should human-robot teams interact to maintain trust, ensure reliance, and reduce stress?''
The problem of autonomous teammates is intensively studied in the context of transparency \cite{[Panganiban-2020]}, level of autonomy \cite{[Patil-2020]}, human-robot interactions (e.g., voice, gesture), and trust \cite{[Lyons-2018]}.  


\subsection{teammates under stress}

\begin{figure}[!ht]
\begin{center}
	\begin{parbox}[h]{0.9\linewidth}{\centering 
			\includegraphics[width=0.48\textwidth]{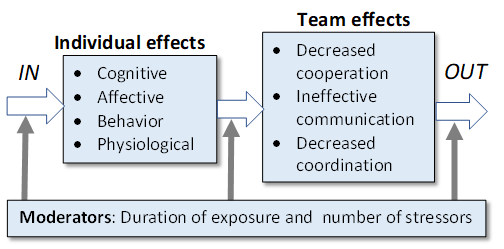}}
	\end{parbox} 
\caption{The stressor (IN) impact on team performance (OUT).}
\label{fig:Stress_Team}
\end{center}
\end{figure}

In \cite{[Dietz-2017]}, team stress has been defined as a ``relationship between the team and its environment, including other team members, that is appraised as taxing or exceeding their resources and/or endangering their well‐being''. Fig. \ref{fig:Stress_Team} illustrates the mechanism of team performance effectiveness (OUT)  under stress (IN)  that  manifests  itself through various effects in both individual processes (cognitive,
affective, physiological, and behavioural) and team processes (cooperation, communication, and coordination) to  impact team performance effectiveness (degree to which products of team effort meets a specific goal related to the criteria of quality or efficiency).  Two focal moderators are incorporated:
1) duration of exposure, and 2) the cumulative, interactive effect of exposure to multiple stressors (e.g., time pressure, task load, threat, uncertainty, fatigues, environmental factors, command pressure, conflict).

\subsection{Propagation paradigm as an exchange of information}

Different phenomena can generally be described and modelled as the exchange of information between members of a population. In \cite{[Vermeer-2018]},  the behaviour propagation, for example,   is defined as a  process by which the behaviour or state
of an agent $A$ will result in a change of behaviour or state between connected neighbours $B$. 
Propagation can be thought as unifying paradigm that  has been discovered and re-discovered in numerous fields under different guises (e.g., 
  rumours \cite{[Daley-1965]},  ideas  \cite{[Bettencourt-2006]},  trust \cite{[Cho-2015]}, risk \cite{[Ojha-2018]}, emotions such as happiness, anger,  sadness, and  fear \cite{[Cornes-2019]},  panic \cite{[Li-2019],[Helbing-2000]}, distress and mental disorder \cite{[Scata-2018]}).
 For example, trust propagation in social networks  can be assessed as trust/distrust (encoded as 0 and 1), risk of decision propagation in terms of high/medium/low risk (encoded as 0, 1, and 2), infection propagation differentiating the susceptible/infected individual (encoded as 0 and 1), and panic propagation in terms of the high-risk-panic/medium-risk-panic/low-risk-panic (encoded by 0,  1, and 2).

Computational epidemiology proposes a network model to describe the spreads of epidemics \cite{[Pare-2020]}. These models can be applied to the spread of e-mail worms and other computer
viruses, the propagation of faults or failures, and, more
generally, the spread of information (e.g., news, rumours, brand
awareness, and marketing of new products), as well as routing in ad-hoc and peer-to-peer networks.

In genomic theory, the network propagation addresses the problem of the network evolving accordingly to the observational data  \cite{[Cowen-2017]}. The evolved networks are Boolean networks for the representation of gene configurations and activity under incorporated randomness \cite{[Karlebach-2013]}. This approach uses the computational logic design methods under the assumption that a gene can be in one of two states.  

We utilize these and other experiences in our development of the stress propagation model. We assume that a combatant can be in a normal state (coded 0) or in a state of stress (coded 1); the  transition between states is based on the Susceptible-Infected-Susceptible (SIS) model \cite{[Martcheva-2015],[Zuzek-2015]}.  

The dynamics of the propagation process were studied in literature from the perspectives of the network structure (path length, clustering, topologies), controllability, predictability, intervention strategies, and under various propagation mechanisms.
For example, authors \cite{[Vermeer-2018]} leverage a perspective from the communication theory to infer the propagation
mechanism. In this approach, the propagation process is   decomposed into three  entities: 1) a sender that encodes
information, 2) a channel (medium) that transports this information from a sender to a receiver, and 3) a receiver that decodes the information. 
While these sub-processes are part the composite propagation
process, each of them is conceptually different from one another.
The models described in  \cite{[Bosse-2015],[Vermeer-2018]} consider the propagation mechanism at the agent level.

Emotion contagion in groups is a social phenomenon. Emotions of a group member can be absorbed by other group members, and can be amplified; in such case, the levels of emotion that may substantially
exceed the original emotion levels of group members may occur \cite{[Bosse-2015]}.



\subsection{Stress propagation in human-robot teams}

 In the study of human-robot teams, the special focus is in the involvement of autonomous robots  as teammates. This is the case when the role of robots is  shifted from the category of an automated tool to the category   of a teammate \cite{[Seeber-2020],[Lawless-2022],[Crandall-2018],[Johnson-2014]}. The key research questions posted, in particular, in \cite{[Rebensky-2022]}, address the following problems: how  the level of autonomy of  machines impact the performance, trust, and effectiveness of human teammate missions and the operator state. The latter includes the ability of the operator to perform target identification, the operator’s stress, workload,  and decision-making in  tactical intelligence surveillance and reconnaissance scenarios.
In particular, it was  reported that team performance, stress, and workload scores indicate that higher levels of autonomy
results in lower levels of stress and workload, and, thus, results in better performance.


 This paper addresses the operation of first-responder teams with autonomous machines as teammates. Specifically, the research question is on the interaction strategies between autonomous machines  and humans  under stressful conditions.

\begin{figure}[!ht]
\begin{center}
\begin{tabular}{c}
	\begin{parbox}[h]{0.9\linewidth}{\centering 
			\includegraphics[width=0.4\textwidth]{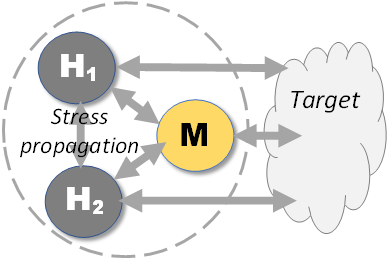}}
	\end{parbox} 
\end{tabular}
\caption{Robot $M$ is an autonomous teammate to humans $h_1$ and $h_2$. Stress  1) can be propagated only between human teammates,  2) can be detected by a robot in human-robot interactions, and 3) can impact the mission. 
}
\label{fig:H-M_Teammate}
\end{center}
\end{figure}

The role of robot as a teammate (instead of role as a tool) is illustrated in Fig. \ref{fig:H-M_Teammate}. Once the  robot detects a stress in a human teammate, it reports to the control and requests to switch to continue the mission with another, unstressed human.

\subsection{Examples of  teammates under stress }

\begin{figure*}[!ht]
	\centering
	\begin{tabular}{cc}
	\begin{parbox}[h]{0.4\linewidth}{\centering 
	\includegraphics[width=0.37\textwidth]{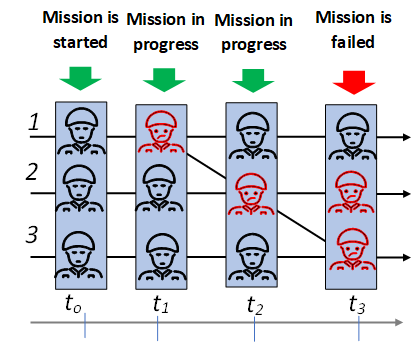}
			}\end{parbox}
			&
	\fbox{\begin{parbox}[h]{0.5\linewidth}{\small
\begin{itemize}
	\item [$-$] At time $t_0$, all combatants are operating in a regular mode. Team status is ``Mission  started''.
	\item [$-$] At time $t_1$, 1st  combatant  is under stressful conditions.   Team status is ``Mission in progress''.
	\item [$-$] At time $t_2$, stress is propagated from 1st to 2nd combatant. however, the 1st combatant is recovered. Team status is ``Mission in progress''.
	\item [$-$] At time $t_3$, stress is propagated from 2nd to 3rd combatant. Team status is ``Mission failed ''.
\end{itemize}
}\end{parbox}}\\				
			\multicolumn{2}{c}{$(a)$}\\
		\begin{parbox}[h]{0.4\linewidth}{\centering \includegraphics[width=0.4\textwidth]{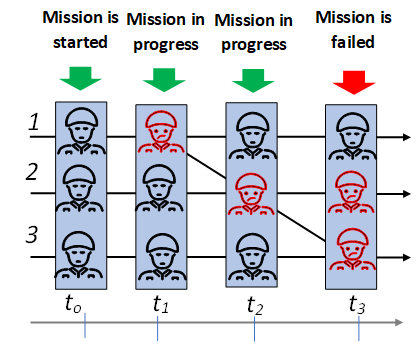}
		}\end{parbox}
		&\
		\fbox{\begin{parbox}[h]{0.5\linewidth}{ \small
\begin{itemize}
	\item [$-$] At time $t_0$, all combatants operate in regular mode. The team status is ``Mission  started''.
	\item [$-$] At time $t_1$, 1st  combatant  is under stress.   Team status is ``Mission  in progress''.
	\item [$-$] At time $t_2$, stress is propagated from the 1st to the 2nd combatant. The robot detects this stress state.  However, the 1st combatant has recovered. The team status is ``Mission in progress''.
		\item [$-$] At time $t_3$, stress is propagated from the 2nd to the 3rd combatant. The robot switches to  cooperation with the 3rd (human) combatant. The team status is ``Mission failed''.
\end{itemize}
}\end{parbox}}\\		
			\multicolumn{2}{c}{$(b)$}
		\end{tabular}
	\caption{Scenarios  of stress propagation in human-only team $(a)$ and human-robot team $(b)$.}
	\label{fig:Stress_Propag_Scenario}
\end{figure*}

Consider two examples  of  stress propagation in teams. 
Given a team of three human combatants,  a potential stress propagation scenario is illustrated in Fig. \ref{fig:Stress_Propag_Scenario}$(a)$.
Given a team of four combatants (three humans and an autonomous machine), there is another scenario as shown in Fig. \ref{fig:Stress_Propag_Scenario}$(b)$. The key difference is that the stress propagation  triggers the process of new team configurations, while the risk of mission failure decreases.

 We define the main requirements to the modelling tools from these simplest scenarios as follows:
\begin{enumerate}
	\item The detailed description of the team configuration and the relationships between the combatants at any time;
		\item  The decision on mission status is updated in real-time.  
			\item  The recovery from stress is a factor of the stress propagation and mission success/failure. 
				\item The defined robot tactic/actions in the case of stress detection.
\end{enumerate}
These requirements describe  the  composite models, such as agent-based models in combination with a specific mechanism for stress propagation (such as SIS) \cite{[Fast-2015],[Gunaratne-2022]}. 

\subsection{Human-machine teammates under stress}

In many practical applications, physiological signals are used to detect the state of human teammates. They are collected using wearable  sensors, often used as a part of the gear for the first-responders  \cite{rodrigues2018wearable}.  They  reliably detect a stress state of a human  \cite{[Lai-2021]}. The stress impacts the human sensory systems such as vision, cognitive processing, and motor skill in  response to  stress,  which are vital for survival.  Under stress, some or all three  systems can break down \cite{grossman2008psychological}, but this is personalized to individuals. 
Recent research roadmaps \cite{Grant2015research,[Panganiban-2020]} highlight the great demand for human-machine teaming. The functions of an autonomous machine in such a team, denoted by $g(M)$, include the detection of affective states of human teammates, including stress. In this paper, the following simplifications are assumed: if $g(M)=0$, the machine operates in a regular mode of teaming. Machine $M$ assigns the  value $g(M)=1$ in communication  with the stressed human.

\section{Problem formulation, 
 approach, and contribution}\label{sec:Framework-approach-problem-formulation}



The overall goal of our study is monitoring first responder missions in stressful environments. The modelling helps  identify high-risk scenarios and  prevent  mission failures, as well as specify requirements and recommendations for combatant training and mission planning. 
The objective of our investigation and modelling is characterized  by  the following  features:

\begin{enumerate}
	\item [$-$] Risk of stress of each teammate is assessed individually using a set of wearable sensors. The stress state can be propagated to other teammates.
	
\item [$-$] The simplest two-state  model represents an unstressed or a stressed teammate.  

	\item [$-$] The first-responder team is relatively small; that is, this team  should be represented by a small-size graph-based model. 
		
		\item [$-$] The relationships between teammates  and levels of their importance is known; they  dynamically  change throughout the mission. 
		
\item [$-$] A decision on mission failure is made if the number of stressed teammates, reaches a certain value critical to the mission execution/failure.
			 
\end{enumerate}

The list of the key features and requirements form the  stress propagation model for a team.  This is  a small-size network where each node corresponds to a teammate. The links between the nodes reflect the relationships between the teammates accordingly to the scenario and the mission. These links can vary throughout the scenario, as well as the cognitive states of  teammates. Thus, the network structure  evolves according to the scenario of the mission. Each scenario consists of a set of scenes.   

In this model, an analogue of the epidemic threshold for contagion must be derived. This is needed to detect a  level of the group stress given a certain time, that is critical or harmful to the mission execution. No additional data is needed for model calibration.

The basic idea behind the proposed stress propagation approach is to combine logical algebra with computational epidemiology models that describe the  propagation of a ''disease'' (or stress in this case) between individuals. This composite model describes the interaction between the teammates, within the scenarios of the mission, team dynamic, and team decision-making. Alternative solutions based on the composition of an agent models and a SIS model offer different possibilities. Specifically, our contribution addresses a model that is
composed of the following advanced  techniques:
\begin{enumerate}
	\item a computational logic that provides a detailed description of the teammate stress states, emphasized on potential configurations (e.g., humans and machines), functionality and task execution; 
		\item a decision-making function that manifests the  state of combat unit at any time;  this function can be changed and updated throughout the mission;
			\item a propagation mechanism adopted from   the  standard SIS model of pandemic propagation. 
\end{enumerate}

The proposed composite model for modelling first responder mission overcomes various limitations of the previously developed models, and provides new opportunities on flexible monitoring of the stress status of combatant teams such as:
\begin{itemize}
	\item [$-$] individually-specific responses on stress and corresponding decision-making,
		\item [$-$] collaborative-specific human-machine relationships depending to the level of machine autonomy, machine strategy depending on the stress factor of human teammates;
\item [$-$] a static, dynamic,  and probabilistic description of a team under stress. 
\end{itemize}

Conceptually, our model is an agent-based  model \cite{[Bonabeau-2002]}. In this approach,  a system is modelled as a
collection of autonomous decision-making entities or agents.
Each agent individually assesses its situation and makes decisions on
the basis of a set of rules. Agents may execute various behaviours
appropriate for the system they represent.  Agent-based  modelling captures emergent phenomena using the interactions of individual entities. 


\section{Methodology}\label{sec:formalization}

In this section,  a framework of our approach is introduced. Three pillars of an approach are distinguished: computational logic,  computational epidemiology, and computational genomics.

\subsection{Computational logic}

The binary nature of stress propagation as well as a wide spectrum techniques for representation, processing, and propagation  data, is the key motivation of focusing on the computational logic and  Boolean data structure. We refer to textbook \cite{[Yanush_Logic_Design_2008]} for details. Specifically, computational logic provides  the following:
\begin{itemize}
	\item [$-$]Ample opportunities for choosing an appropriate data structure. For example, a truth table and/ or a logical expression (e.g., sum of products, logic and arithmetic polynomials) are well suited  forms for expressing the  stress scenarios for small teams.
		\item [$-$]A scalable solution in the form of a graph-based data structure called the Decision Diagram (DD). There are many guides  on the DD construction and application, e.g.,  \cite{[Yanush_Decision_Diagram_2006]}.
	 The DD is well-suited for detailed stress propagation modelling in both small and large  teams. 
			\item [$-$]A convenient uncertainty representation. A logic data structure can convey uncertainty, or ``don't cares values", i.e., the logic values (of 0, 1 or more in case of multi-valued logic) can be replaced by the probability of their occurrence.  The ``stress'' or ``no-stress'' (0 or 1) state can be represented by their probabilities, e.g., the stress in one combatant is detected with a probability 0.7, and 0.8 in the second one. This area is called a ``probabilistic logic computing'' \cite{[Yanush_Noise_Resilient_2013]}. 
\item [$-$]A propagation mechanism. For example, fault propagation  is a common practice in computational logic.  There are various similarities between stress propagation in a team and fault propagation in a logic network. 
\item [$-$]Tools for dynamic and time-domain modelling. Computational logic provides possibilities for modelling dynamics in various ways, for example, using Boolean differential calculus and evolving data structures \cite{[Yanush_Decision_Diagram_2006]}. 
\end{itemize}


\subsection{Computational epidemiology}

Computational epidemiology focuses on modelling the propagation dynamic. The similarities of infection and stress propagation are based on the following.  
The group of short-term immunity memory diseases (or population) is represented by the  SIS model \cite{[Martcheva-2015]}. The SIS model is one from a collection of computational epidemiological models of disease spread that form the standard toolkit of various modifications and applications.  The SIS model supposes a Susceptible $(S)$ state with  the probability  $\beta$, i.e.,  individuals who are healthy but can contract the disease;  and Infected  $(I)$ with  the probability $\gamma$, i.e., individuals who have contracted the disease and are now sick with it. Individuals randomly go through the cycle of
$S \leftrightarrow I $. Similarities with stress propagation are obvious if the disease is replaced by stress. Previously, we noted that these similarities are often utilized for various technology-societal categories such as trust, risk, ideas, rumours, panic, etc.   The following formalism is useful for the purposes of a stress propagation model (partially used in our approach):
\begin{itemize}
	\item [$-$]Dynamics of  class  $(I)$ are described by the differential equation $I^{'}=\alpha \times I$  with initial condition at time zero $I(0)=I_0$.
		\item [$-$]Solution is the number of people in the infectious
class at time $t$ is given by  $I(t)= I_0\times e^{-\alpha \times t }$.
			\item [$-$]From this follows that the  probability of recovering/leaving the infectious class is calculated from cumulative probability distribution  $F(t)=1-e^{-\alpha \times t }, t\geq 0$, or probability density function $f(t)=\alpha \times e^{-\alpha \times t}$.
\item [$-$]Mean time spent in the infectious class is $1/\alpha$. Recovery rate can be calculated if the mean time is known. 
\end{itemize}
   For example, in the case of 5 days influenza, the recovery rate, measured in units of [days]$^{-1}$, is 1/5. In the case of stress propagation, days can be replaced by seconds.
For details we refer the text \cite{[Martcheva-2015]}.

Composite models based on SIS paradigm are well known.
For example, computational logic was used in \cite{[Chen-2017]} to extend the  SIS model  for modelling single contagion processes.


\subsection{Computational genomics}

The fundamental tenet of computational genomics is the network propagation using the discovered data \cite{[Karlebach-2013],[Li-2018],[Leifeld-2018],[Zobolas-2022]}. This is  
a system identification problem aiming at building the mathematical models of dynamical systems from measured data.  
The network propagation transforms a short list of candidate genes into a genome-wide profile of gene scores. 
This transformation greatly improves the power of genetic association, providing a universal amplifier for genetic structure. Overview \cite{[Schwab-2020]} provides advances in probabilistic Boolean networks for dynamic behaviour of biological systems.

Computational methods of genomics focuses on the discovery of gene network configurations using computational logic (Boolean networks) and belief propagation techniques such as Bayesian network factor graphs and Markov chains.


\subsection{Definitions}

 This section presents the basic preliminaries of current work along with an adaptation of logic data structures for specific needs of stress propagation.

\begin{figure*}[!h]
\begin{center}
\begin{small}
\renewcommand{\arraystretch}{1.3}
\begin{tabular}{cc}
   \begin{parbox}[h]{0.55\linewidth} {\centering
	\renewcommand{\arraystretch}{1.3}
  \begin{tabular}{|cc||cl||l|}
	    \hline
	\multicolumn{2}{|c||}{\bfseries  teammate} & \multicolumn{2}{c}{\bfseries State}   &\multicolumn{1}{||c|}{\bfseries  Decision} \\
\multicolumn{2}{|c||}{\bfseries  Scene} & \multicolumn{2}{c}{\bfseries of Combat Unit} & \multicolumn{1}{||c|}{\bfseries on  Mission}\\
\mbox{\small $h_1$} & \mbox{\small $h_2$}&
 \multicolumn{2}{c}{$f(h_1,h_2)$}& \multicolumn{1}{||c|}{}
        \\
    \hline
0 & 0  & $\mathbf{0}$ &$\Rightarrow$ Regular mode & {No action}\\
0 & 1  & $\mathbf{1}$ &$\Rightarrow h_2$ is stressed    &{No action}\\
1 & 0  & $\mathbf{1}$ &$\Rightarrow h_1$ is stressed   &{No action}\\
1 & 1  & $\mathbf{X}$ &$\Rightarrow$ Uncertain    &{Uncertain}\\
 \hline
\end{tabular}}
 \end{parbox}& 
  \begin{parbox}[h]{0.3\linewidth} {\centering
	\includegraphics[width=0.25\textwidth]{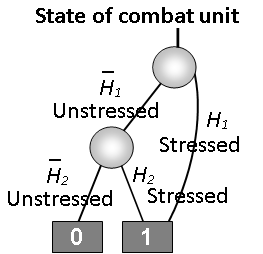}
	}
 \end{parbox}\\\\
\multicolumn{1}{c}{$(a)$}& \multicolumn{1}{c}{$(b)$}
\end{tabular}
\end{small}
		\end{center}
\caption{State of combat unit of two combatants $h_1$ and $h_2$: Decision table   $(a)$   and  DD $(b)$.
  }\label{fig:SIS}
\end{figure*}

\begin{defn}
 Given a teammate $h_i,i=1,2,\ldots , n$ that can be in an operational (unstressed) state or stressed state, encoded by 0 and 1, respectively, \textbf{Tabulated state}  of teammate is defined as a specification of  $h_i=\sigma, \sigma_i \in \{0,1\}$.
Given $n$ teammates $h_1,h_2,\ldots ,h_n$, their tabulated states are defined as  $h_1h_2\ldots h_n=\sigma_1\sigma_2\ldots \sigma_n$.
\end{defn}
Example of tabulated states of two teammates $h_1$ and $h_2$
 is given in Fig. \ref{fig:SIS}a: $h_1h_2=\{00\},\{01\},\{10\},\{11\}$.

\begin{defn}
\textbf{Stress response function} $g(m)\in\{0,1\}$ of an autonomous machine $m$ is defined as the manifestation by ``1''   detection stressed human teammate vs. ``0'' as a regular mode.
\end{defn}

\begin{defn}
Given $i+j$ teammates: humans $h_i,i=1,2,\ldots , n$  and autonomous machines $m_j, j=1,2,\dots , k$. \textbf{State of combatant unit} is a function 
$f(h_i,g(m_j))$ where $g(m_j)$ is the  stress response function of machine $m_j$.
 This function is evolved with respect to the teammate states.
\end{defn}

Examples for two human teammates $h_1$ and $h_2$ are given in Table \ref{tab:State-combatant-unit} that represents four of $2^4=16$ possible scenarios. The first  scenario   addresses the request of a decision-making system  to detect stress in a team. The second scenario  detects causal relationships between the combatants and their stress. The third scenario  corresponds to the recovering situation, i.e.,  risk that both combatants will be  stressed is negligible. The fourth scenario describes regular team functioning.
There are other scenarios which are not included in the table, e.g., $\overline{h}_1\vee h_2, {h}_1\vee \overline{h}_2, \overline{h}_1\wedge \overline{h}_2, $ etc.
This dynamic of state of the team is conceptually similar to the network propagation in genomic theory.

\begin{table}[!h]
\renewcommand{\arraystretch}{1.3}
\caption{Notion of the state of a combatant unit $f(h_1,h_2)$ in a set of scenarios.}\label{tab:State-combatant-unit}
\begin{center}
  \begin{tabular}{|c|l|c|}
		 \hline
	\bfseries Scenario & \bfseries State of combatant unit & \bfseries Formalization\\

1&
\begin{parbox}[h]{0.38\linewidth} {
 \vspace{1mm} At least one of the combatants is stressed\vspace{1mm}
 }\end{parbox}
&
\begin{parbox}[h]{0.33\linewidth} {
$h_1\vee h_2=[0111]^T$
 }\end{parbox}\\
 \hline
2&
\begin{parbox}[h]{0.38\linewidth} {
 \vspace{1mm}One of the combatants is stressed\vspace{1mm}
 }\end{parbox}
&
\begin{parbox}[h]{0.33\linewidth} {
$h_1\oplus h_2=[0110]^T$
 }\end{parbox}\\
 \hline
3&
\begin{parbox}[h]{0.38\linewidth} {
 \vspace{1mm}Both combatants are stressed\vspace{1mm}
 }\end{parbox}
&
\begin{parbox}[h]{0.33\linewidth} {
$h_1\wedge h_2=[0001]^T$
 }\end{parbox}\\
 \hline

4&
\begin{parbox}[h]{0.38\linewidth} {\vspace{1mm}
Both combatants operate
 }\end{parbox}
&
\begin{parbox}[h]{0.33\linewidth} {\vspace{1mm}
$\overline{h_1\vee h_2}=[1000]^T$ 
 }\end{parbox}\\
 \hline
\end{tabular}   
	\end{center}
\end{table}

\begin{defn}
\textbf{Decision table} is a composite data structure that conveys three kinds of information:
1) the tabulated states of teammates;
2) the tabulated states of combat union; and
3) a decision on the mission.
\end{defn}

Example of a decision table is given in Fig. \ref{fig:SIS}a.
Consider two combatants, $h_1,h_2\in \{0,1\}$, where unstressed and stressed combatants are encoded as  0 and 1, respectively. 
The possible states of teammate $h_1$ and $h_2$ are tabulated, they  produce the $2^2=4$ state space. The state of the combat unit is represented by the tabulated function $f(h_1,h_2)=[0111]^T=h_1\vee h_2$ and a comment. 

A decision on the mission, i.e., actions supporting the task may be missed, delayed, or failed if a teammate's affective state is specified as stressed. 

The human-robot teams are groups of at least two teammates, working toward a common goal. An example of decision table of  two teammates, human $h$ and machine $m$ is given in  Fig. \ref{fig:SIS-h-M}$(a)$. A machine $m$ manifests $g(m)=1$ if that action is required if a stressed human teammate $h$  is identified. 
Formally, the state of a combat unit is described by $f(h,g(m))=h\vee g(m)$. The ignored assignments $\{01,10\}$ are called "don't cares". 

Fig. \ref{fig:SIS-h-M}$(b)$ provides additional details about the three members of the team: two humans, $h_1$ and $h_2$, and an autonomous  machine $m$. 
 Communication between a machine $M$  and a stressed human $h_1$ or $h_2$ results in the blockage of the stress propagation, while the mission  continues. However, the situation is uncertain when one of stressed humans can be recovered, i.e., $f(h_1,h_2,g(m))=g(m)(h_1\oplus h_2)$.     

 A decision is characterized as risky when the expected outcome is uncertain, the goals of the decision are difficult to achieve, and/or  the possible outcomes could have extreme consequences \cite{[Hagemann-2022]}. The decision table with "don't cares" formalizes this notion.

The tabulated form of function $f$ of the state of a combatant unit is good for illustrative purposes but unpractical, because its size growth is exponential. This is the main reason to use scalable graphical data structures such as decision diagrams.
\begin{defn}
\textbf{Decision Diagram} (DD) is a graphical data structure as a directed acyclic graph  for representation and computing logic expressions. The DD consists of two kinds of nodes: a) operational nodes and b) terminal nodes.   In DD, logical expressions are  represented  by a unique paths. 
\end{defn}
In the case of binary logic, the DD has a 0-terminal and a 1-terminal node. Every non-terminal (operational) node  has an index to identify an input variable of the Boolean function, and has two outgoing edges, called the 0-edge and 1-edge.

Example is given in Fig. \ref{fig:SIS}b. This    DD is a graphical representation with an incorporated size-reducing mechanism. For Boolean functions, the number of terminal nodes is equal to two: 0 and 1. The number of operational nodes with Shannon expansion depends on the complexity of Boolean functions. For this function, only two nodes are needed to represent $2^2=4$ states. Three paths of this DD represent the decision table:    
\begin{eqnarray*}
&\text{Path 1:}~&\overline{h}_1\rightarrow \overline{h}_2\rightarrow \fbox{0}\equiv h_1h_2=\{00\};\\
&\text{Path 2:}~&\overline{h}_1\rightarrow {h}_2\rightarrow \fbox{1}\equiv h_1h_2=\{01\};\\
&\text{Path 3:}~&{h}_1\rightarrow \fbox{1}\equiv h_1h_2=\{10,11\}
\end{eqnarray*}
Details of DD can be found  in \cite{[Yanush_Decision_Diagram_2006]}.

\begin{figure*}[!h]
\begin{center}
\begin{small}
\renewcommand{\arraystretch}{1.3}
\begin{tabular}{cc}
   \begin{parbox}[h]{0.45\linewidth} {\centering
	\renewcommand{\arraystretch}{1.3}
 \begin{tabular}{|cc||cl||l|}
	    \hline
	\multicolumn{2}{|c||}{\bfseries  Teammate} & \multicolumn{2}{c}{\bfseries State}   &\multicolumn{1}{||c|}{\bfseries  Decision} \\
\multicolumn{2}{|c||}{\bfseries  Scene} & \multicolumn{2}{c}{\bfseries of Combat Unit} & \multicolumn{1}{||c|}{\bfseries on  Mission}\\
\mbox{\small $h$} & \mbox{\small $g(m)$}&
        \multicolumn{2}{c}{\mbox{\small \text{$f(h,g(m),t)$}}}& 
        \multicolumn{1}{||c|}{}\\
    \hline
0 & 0  & $\mathbf{0}$ &$\Rightarrow$ Operate  &\text{No action}\\
1 & 1  & $\mathbf{1}$ &$\Rightarrow h$ is stressed    &\text{Action}\\
 \hline
\end{tabular}}
 \end{parbox}& 
  \begin{parbox}[h]{0.5\linewidth} {\centering
	\renewcommand{\arraystretch}{1.3}
  \begin{tabular}{|ccc||cl||l|}
	\hline
	\multicolumn{3}{|c||}{\bfseries  Teammate} & \multicolumn{2}{c}{\bfseries State}   &\multicolumn{1}{||c|}{\bfseries  Decision} \\
\multicolumn{3}{|c||}{\bfseries  Scene} & \multicolumn{2}{c}{\bfseries of Combat Unit} & \multicolumn{1}{||c|}{\bfseries on  Mission}\\
\mbox{\small $h_1$} & \mbox{\small $h_2$}& \mbox{\small $g(m)$}&
        \multicolumn{2}{c}{\mbox{\small \text{$f(h_1,h_2, g(m))$}}}& 
        \multicolumn{1}{||c|}{}\\
    \hline
0 &  0  &0  &$\mathbf{0}$ &$\Rightarrow$  Operate  &\text{No action}\\
0 & 1  &1  & $\mathbf{1}$ &$\Rightarrow h_2$ is stressed     &\text{No action}\\
1 & 0  &1  & $\mathbf{1}$ &$\Rightarrow h_1$ is stressed     &\text{No action}\\
1 & 1  &1  & $\mathbf{X}$ &$\Rightarrow$ Uncertain    &\text{
Uncertain}\\
 \hline
\end{tabular}}
 \end{parbox}\\\\
\multicolumn{1}{c}{$(a)$}& \multicolumn{1}{c}{$(b)$}
\end{tabular}
\end{small}
		\end{center}
\caption{
Stress propagation in human-machine team: $(a)$  machine $m$  recognizes the stress state of human $h$ and manifests that action as needed;   $(b)$ machine $M$  recognizes the stress state of human $h_1$ and $h_2$ and manifests that action as needed only if both human teammates are stressed.
  }\label{fig:SIS-h-M}
\end{figure*}

In this paper, a team of combatants under stress is considered as a binary system. Change of state (stressed or not) of a single or group of combatants should be detected.  For this, Boolean difference can be adopted.


\begin{defn}
\textbf{Factor graph} is  an undirected (bipartite)  graph representing the factorization of a multivariate function into a product of functions (factors) \cite{[Koller-2009]}.
\end{defn}

The factor graph representation provides a visual tool
for the decomposition of a high-dimensional joint posterior
distribution into smaller, tractable parts.
  In probability theory and its applications, factor graphs are used to represent factorization of a probability distribution function, enabling efficient computations, such as the computation of marginal distributions through the message-passing algorithm \cite{[Kschischang-2001]}. This algorithm is a general rule in factor graphs that
allows to iteratively approximate marginal (posterior) distributions.

 Factor graphs contain two types of nodes: one type corresponds to random variables; the others corresponds to factors over variables. The graph only contains edges between variable nodes and factor nodes. Factor graphs were  extended to encompass both Bayesian and Markov networks  \cite{[Frey-2003]}.
Detailed study of factor graphs and message passing algorithm are introduced in \cite{[Koller-2009]}.

The key goal of simulation for stress propagation in teams is to  estimate the probability that a given combatant will become stressed at a
particular time as a result of stress propagation from another combatant. These marginals help formulate  recommendations for training combatants and mission planing such as determining an optimal number  of combatants and level of trust with robot teammates \cite{[Briggs-2021],[Lawless-2022],[Lin-2022],[Lyons-2018],[Johnson-2014]}, minimizing the risk of stressed combatant, and  the probability that a team will remain unstressed after a fixed time \cite{[Gamble-2018]}.

Monte-Carlo experiments provide answers to these questions but it is computationally expensive because  many independent trials are needed in order to collect satisfactory statistic data. Message-passing method overcomes these limitations. 
Stress (emotion) propagation similar to infection propagation  can be simulated using Bayesian network \cite{[Urizar-2016],[Xiong-2021]}, epidemiological model using differential equations,  e.g., SIS and  SIR \cite{[Martcheva-2015]}, message-passing algorithm \cite{[Koller-2009],[Kschischang-2001]}, as well as composite model  \cite{[Shrestha-2015]} in which differential
equations are used for the messages over time.
 Propagation using differential equations for the messages have been proposed in \cite{[Shrestha-2015]}.

\section{Composite stress propagation model}
\label{sec:model}

From a computational perspective,  the proposed model combines two kinds of graphical data structures: DDs (their nodes represent logical operations) and  stress propagation graphs (their nodes represent agents)


\subsection{Composite techniques}

The spread of stress depends on both the amount of contact between teammates
and the possibility of an stressed person transmitting stress to another teammate. If each stressed combatant is in contact with two other combatants before they recover, the stress will soon begin to spread rapidly. Assuming that recovery requires one minute, the number of stressed combatants will double each minute.

In \cite{[Fast-2015]}, a composite model was developed using  the joint diffusion of social
response and disease through a population. Authors implemented a SIR model,
adapted for agent-based modelling. Specifically, in addition  to the disease state, each agent has a value
associated with his social response.
Agent-based models using probabilistic cellular automata to simulate SIR  dynamics using COVID-19 infection parameters have been implemented in \cite{[Lima-2021]}.
Work \cite{[Lazebnik-2022]} extends the traditional SIR model by introducing  exposed groups, asymptomatic and symptomatic infected people, and  dead group, including spatio-temporal interactions regarding different buildings, social contexts, and countermeasures (isolation, distancing, mask wearing, vaccination).
 Spatial agent-based SIR models
to evaluate the relative risk reduction of intervention strategies was introduced in \cite{[Gunaratne-2022]}. Intervention strategies were decomposed into three dimensions, namely movement restriction, population
capacity, hourly break probability.
Researchers  \cite{[Paoluzzi-2021]} developed
a single-agent extension of the SIR model where mobility is taken into account. In addition to the infection and removing rate, the
mobility factor is added.


\subsection{Proposed model}

\begin{figure}[!ht]
\begin{center}
			\includegraphics[width=0.5\textwidth]{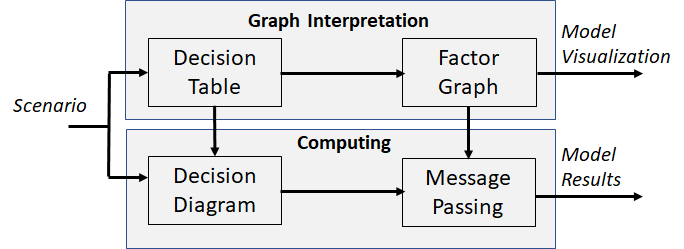}
\caption{Composite model of stress propagation in first responder team.
}
\label{fig:Model_Composite}
\end{center}
\end{figure}

A composite model consists the graph interpretation part and   computational part over binary or multi-valued  data structure  (Fig. \ref{fig:Model_Composite}):
\begin{itemize}
	\item [$-$]  The graph interpretation part provides visual interpretation of a given part of scenario using decision table (graph) and factor graph.
		\item [$-$] The computational part includes decision diagram that represents overall scenario and message passing algorithm for stress propagation.
			\item [$-$]  The data structure  consisting technique for description scenario in terms of two-valued (Boolean) logic or multiple-valued logic.
\end{itemize}

Note that the decision table (graph) and the DD carry out a representation of potential scenarios  and enable reasoning over the deterministic variables. The factor graph and message passing algorithm provide mechanism to reason probabilistically about the values of one or more of the variables. In order to do so, a joint distribution over space of possible assignments to some set of random variables. This part of model allows to answer a broad  range of queries using factorization and marginalization mechanisms, i.e., break up the joint distribution into smaller factors, each over a smaller space of possibilities; as well as  define the overall joint distribution  as a product of these factors. Factors can be viewed as the ``compatibilities'' between values of the variables.

\begin{table}[ht]
\renewcommand{\arraystretch}{1.3}
\begin{center}
  \caption{Sample of benchmarks  for modelling stress propagation in first responder teams.}\label{tab:benchmark}
  \begin{small}
\begin{tabular}{|lllll|}\\
\hline
\bfseries Name & \bfseries Teammate & \bfseries Scenario   & \bfseries Node & \bfseries Path\\
    \hline
dc1       &  4      & 7     &22   &$7\times 2^4$       \\
rd53      &  5      & 3     &21   &$3\times 2^5$       \\
sgr6      &  6      & 12    &127   &$12\times 2^6$       \\
sqn       &  7      & 3     &87   &$3\times 2^7$       \\
adr4      &  8      & 5     &123  & $5\times 2^8$  \\
9sym      &  9      & 1     &35   &$1\times 2^9$        \\
sym10     &  10     & 1     &40 &$1\times 2^{10}$      \\
alu1      &  12     & 8     &1003 &$8\times 2^{12}$      \\
co14      &  14      & 1     &29   &$1\times 2^{14}$       \\
gary       &  15     & 11     &510   &$11\times 2^{15}$       \\
in1       &  16      & 17     &617   &$17\times 2^{16}$       \\
opa       &  17      & 69     &313   &$69\times 2^{17}$     \\
vg2       &  25      &  8    &1100 &$8\times 2^{25}$        \\
\hline
\end{tabular}
 \end{small}
 \end{center}
\end{table}

\section{Computational guidelines }\label{sec:guidelines}

In this section, the computational guidelines for stress propagation based on the proposed composite model (Fig. \ref{fig:Model_Composite})  are introduced. The guidelines include the following steps: 

\begin{enumerate}
	\item [] Step I: Goals  and task formalization;
		\item [] Step II: Deriving the decision table;
				\item [] Step III: Deriving the decision diagram;
						\item [] Step IV: Deriving factor graph;
\item [] Step V: Inference and reasoning for decision support.
\end{enumerate}

\subsection{Framework example}

As a framework for this guidelines, the following combat  scenario is considered.

$-$ Consider a group of first-responders consisting of four teammates: three humans $h_1,h_2,h_3$  and autonomous machine $m$ with the stress respond function $g(m)$,

 $-$ A mission  consists of a set of scenarios and each of them includes a number of scenes.	One of scenes is given in Fig. \ref{fig:4-combat} where cooperation with robot $m$ is the prerogative of human teammate $h_2$ and $h_1$ can collaborate  with $h_3$ only through $h_2$, that is, no way to propagate stress from $h_1$ to $h_2$. 
 
$-$ Mission  fails if two human teammates are failed (stressed);
			
 $-$ Teammates $h_1,h_2,h_3,g(m)$ are assigned their factors $f_1,f_2,f_3,f_m$ respectively.

\subsection{Step I: Goals  and task formulation}

The goal of stress propagation modelling is to assess the effectiveness of human-teammate missions and operator state in various (including rare) potential  scenarios.  

A sample of tasks from the  list of modelling tasks  in order to do training, planning, and management improvements includes:

\begin{itemize}
	\item [$a)$]Team's and individual teammates' performance in various stressed scenarios and particular scenes;
		\item [$b)$]Team performance with respect to recovering from a stressed state; 
			\item [$c)$]Detection of behavioural biases  related to the stress state of a human teammate;
				\item [$d)$]Impacts of the level of machine autonomy in team performance;
					\item [$e)$]Impact of trust between human and machine teammates into mission planning and control. 
\end{itemize}

\subsection{Step II: Deriving the decision table}

\begin{figure*}[!h]
\begin{center}
\begin{small}
\begin{tabular}{cc}
    \begin{parbox}[h]{0.3\linewidth} {\centering
\includegraphics[width=0.26\textwidth]{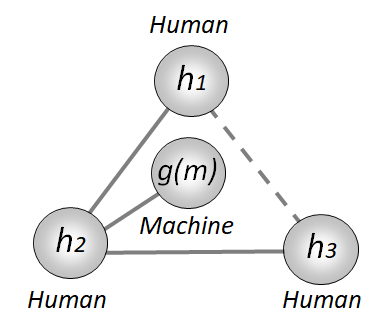}
    }\end{parbox}
		&
    \begin{parbox}[h]{0.6\linewidth} {\centering
\renewcommand{\arraystretch}{1.3}
\begin{tabular}{|cccc||cl||cl|}
\hline
 \multicolumn{4}{c||}{\bfseries Teammate scene}&\multicolumn{2}{c||}{\bfseries State of Combat Unit}&\multicolumn{2}{c|}{\bfseries Decision}\\
\mbox{\small $h_1$} & \mbox{\small $h_2$}& \mbox{\small $h_3$}&\mbox{\small $g(m)$}&
\multicolumn{2}{c||}{ $f(h_1,h_2,g(m))$}& 
\multicolumn{2}{c|}{\bfseries on Mission}\\
    \hline
0 & 0& 0  & 0  & $\mathbf{0}$ &$\Rightarrow$ Regular mode                  &  &\text{No action}\\
0 & 0 & 1  & 0  & $\mathbf{0}$ &$\Rightarrow h_3$ is stressed     &  &\text{No action}\\
0 & 1 & 0  & 1  & $\mathbf{0}$ &$\Rightarrow h_2$ is stressed   & &\text{No action}\\
0 & 1 & 1  & 1  & $\mathbf{1}$ &$\Rightarrow h_2$ and $h_3$ are stressed   &  &\text{Mission  fails}\\
1 & 0 & 0  & 0  & $\mathbf{0}$ &$\Rightarrow h_1$ is stressed                   &  &\text{No action}\\
1 & 0 & 1  & 0  & $\mathbf{1}$ &$\Rightarrow h_1$ and $h_3$ are stressed     &  &\text{Mission  fails}\\
1 & 1 & 0  & 1  & $\mathbf{1}$ &$\Rightarrow h_1$ and $h_2$ are stressed   & &\text{Mission  fails}\\
1 & 1 & 1  & 1  & $\mathbf{1}$ &$\Rightarrow h_1,h_2$, and $h_3$ are stressed   &  &\text{Mission  fails}\\
 \hline
\end{tabular}}
 \end{parbox}\\\\
$(a)$&$(b)$
\end{tabular}
\end{small}
\caption{$(a)$ Graphical representation of combat scenario of  human teammates $h_1,h_2,h_3$ and robot teammate $m$; robot collaborates only with human teammate $h_2$;  $(b)$ a decision table of this scenario as a set of scenes.  }\label{fig:4-combat}
		\end{center}
\end{figure*}

A decision table in Fig. \ref{fig:4-combat}b is derived accordingly to the scenario   given in Fig. \ref{fig:4-combat}a. Human teammates' affective states are represented by all $2^3=8$combinations of their possible  states. Robot-teammate $m$ with  stress respond function $g(m)$  is able to detect stress of human $h_2$: stress is detected using the stress respond function:   if  human teammate $h_2=1$ is stressed then $g(m)=1$. State of combat union manifests ``1'', i.e., that mission is failed, only if at least two human-combatants are stressed. This decision table contains $2^3=16$ assessments with "don't care" outcomes (Fig. \ref{fig:4-combat}b).

\subsection{Step III: Deriving the  decision diagram}

\begin{figure*}[ht]
\begin{center}
\begin{tabular}{ccc}
 \begin{parbox}[h]{0.4\linewidth} {\centering
\includegraphics[scale=0.6]{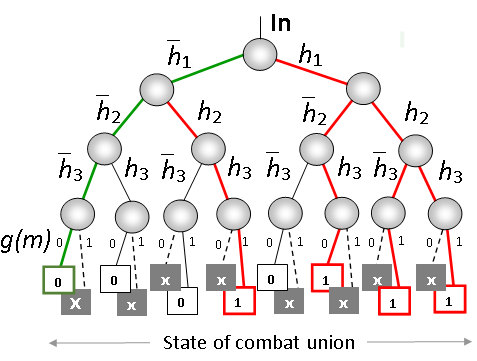}
}
\end{parbox}&
\begin{parbox}[h]{0.12\linewidth} {\centering
\includegraphics[scale=0.7]{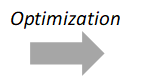} 
}\end{parbox}
&
 \begin{parbox}[h]{0.3\linewidth} {\centering
\includegraphics[scale=0.6]{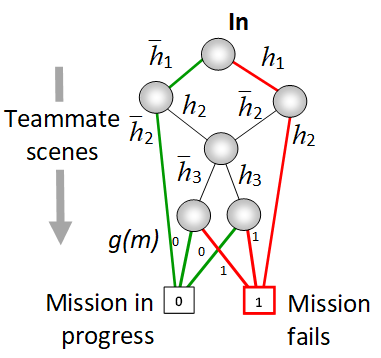} 
}\end{parbox}
 \end{tabular}
\end{center}
\caption{$(a)$ Decision tree of four combatants accordingly decision table in Fig. \ref{fig:4-combat}b. Each path corresponds the scene of combatant scenario. $(b)$ Decision diagram as optimized decision tree.}\label{fig:Tree-4-combatants}
\end{figure*}

Using the decision table (Fig. \ref{fig:4-combat}b), a decision tree in Fig. \ref{fig:Tree-4-combatants} is derived as follows.

 $-$ There are $2^4=16$ states of combat union of four combatants but only half of them are specified. Other states are "don't care" states denoted by `X'.

 $-$ Each node of the decision tree represents  Shannon expansion that decomposed Boolean functions with respect to given variable, i.e.,  $\overline{h}_i$ and $h_i$, and $\overline{g}(m)=0$ and $g(m)=1$. 

 $-$ There are 8 paths in the tree, each path corresponds to the scene of scenario. For instance, path $\overline{h}_1\overline{h}_2\overline{h}_3\overline{g}(m)=0$ (green) corresponds the scene $h_1h_2h_3g(m)=0000$ that results the 0-state of combat union.

$-$ Failure paths (red) lead to the terminal nodes ``1''. For example, the last line in the decision table corresponds to the failure scene $h_1h_2h_3g(m)=0000$ and is represented by the red path $h_1,h_2,h_3,g(m)=1$.

\subsection{Step IV: Deriving the factor graph}

Among various probabilistic models  (e.g., Bayesian networks, Markov random fields), factor graphs are chosen to model stress propagation in team.
In practice, the information available for stress propagation mechanisms
is incomplete and of variable certainty, which motivates the
employment of a probabilistic framework for inferring such propagation functions.

 Each stress-related entity can be modelled as a discrete or continuous random variable.
This random variable represents the
level in which an entity is present in the stress propagation process. Taking partial information
into account, a probabilistic distribution function can be derived  for each variable
that considers interaction relations among them as well as their level
of uncertainty.

\begin{figure}[!ht]
\begin{center}
\begin{tabular}{cc}
 \begin{parbox}[h]{0.4\linewidth} {\centering
\includegraphics[scale=0.4]{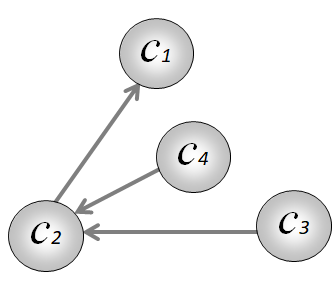}
}
\end{parbox}&
\begin{parbox}[h]{0.42\linewidth} {\centering
\includegraphics[width=0.24\textwidth]{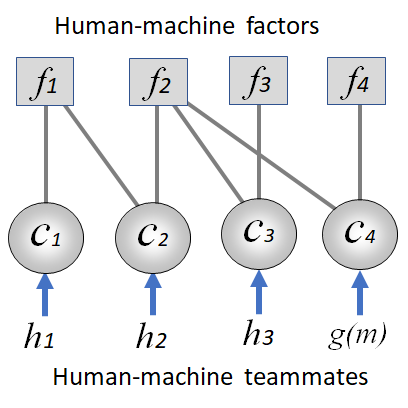}
}\end{parbox}\\
$(a)$&$(b)$
\end{tabular}
\caption{Graphical representation of stress propagation scenario for three human teammates and machine-teammate. $(a)$ Bayesian network $(b)$ Factor graph.}
\label{fig:MessagePas_4_nodes}
\end{center}
\end{figure}

 Consider a set of
discrete random variables (nodes) $c_1=h_1,c_2=h_2,c_3=h_3,c_4=g(m)$. Each node 
may take a value from a range of (usually finite) $M$ logical states. In this paper, we consider only binary case.
Let us denote $Pa_i=Pa_{i1},Pa_{i2},Pa_{i3}, Pa_{i4}$ the set of \emph{parents} of variable   $p(c_i)$. 
 By definition, the factor function $f_i$ 
for the $i-$ teammate is formulated by the conditional probabilities as 
\begin{eqnarray*}
f_{i} &\stackrel{\text{def}}{=}& p(c_i)|Pa_i)
\end{eqnarray*}
Function $f_i$ are referred to as the belief that $f_i$ takes a certain
state with respect to an assignment from its parents. If there are no parents, then  $p(c_i|\oslash=p(c_i)$. 
These formalization results Bayesian network (Fig. \ref{fig:MessagePas_4_nodes}a) 
\begin{eqnarray*}
  p(c_1,c_2,c_3,c_4)&=&\prod_{i=1}^{4}p(c_i)|Pa_i)
\end{eqnarray*}
and factor graph (Fig. \ref{fig:MessagePas_4_nodes}b) 
\begin{eqnarray*}
 p(c_1,c_2,c_3,c_4)&\propto& p(c_1|c_2)p(c_2|c_3,c_4)p(c_3)p(c_4)
\end{eqnarray*}
To convert a Bayesian network to a
factor graph, simply draw an edge between a variable $c_i$  and a factor
$f_j$, if the scope of $f_j$ contains $c_i$.

\subsection{Step V: Inference and reasoning for decision support}

\begin{figure}[!h]
\begin{center}
\includegraphics[scale=0.4]{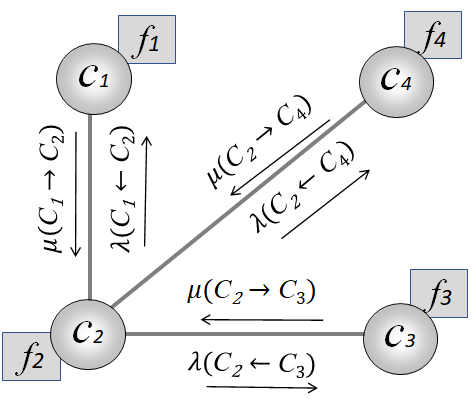}
\caption{Message-passing algorithm.}
\label{fig:Message-Passing-Algorithm}
\end{center}
\end{figure}

In many circumstances, we seek to compute the posterior
distributions, also referred to as marginal functions, $p_i(c_i)$.
Accordingly message-passing protocol, belief propagation between  nodes is implemented by two kinds of messages: from parent to child and from child to parent. To illustrate this propagation protocol, factor graph from Fig. \ref{fig:MessagePas_4_nodes}b  is slightly modified as shown in   Fig. \ref{fig:Message-Passing-Algorithm}, where a message sent 

\begin{itemize}
	\item [] \hspace{-9mm}$-$ from  $c_1$ to $c_2$ and vice versa  is denoted as $\mu(c_1\rightarrow c_2)$ and $\lambda (c_1\leftarrow c_2)$, respectively;
	
	\item []\hspace{-9mm}$-$  from  $c_2$ to $c_3$ and vice versa is denoted  as $\mu(c_2\rightarrow c_3)$ and $\lambda (c_3\leftarrow c_2)$, respectively;
	
	\item []\hspace{-9mm}$-$  from  $c_2$ to $c_4$ and vice versa is denoted  as $\mu(c_2\rightarrow c_4)$ and $\lambda (c_4\leftarrow c_2)$, respectively.
\end{itemize}

\subsection{Recommendations on software tools}

All parts of the proposed composite models are supported by publicly available software packages and benchmarks, in particular, 
\begin{enumerate}
	\item The DD packages is available from \url{https://github.com/johnyf/tool\lists/blob/main/bdd.md}
		\item The message-passing, factor graph, Bayesian Network and related libraries are available from \url{https://pgmpy.org/models/factorgraph.html}
		\item The Bayesian network package is also available from \url{https://pyagrum.readthedocs.io/en/1.3.2/}.
			\item Benchmarks are available from \url{https://github.com/lsils/benchmarks}.
\end{enumerate}

A benchmark can be viewed  as a standard against which the performance of different approaches can be measured. Benchmarking is a process where different approaches can be compared by  identifying  gaps in performance. Benchmarks should satisfy a number of requirements such as transparency, scalability, representativeness, repeatability, and cost-effectiveness. Collection of LGSynth91 benchmarks \cite{[Yang-1989]} satisfies these requirements, it is the ``gold standard'' in computational logic. This collection is applicable in modelling stress propagation because this paper proposes problem formulation in terms of computational logic. Note that the LGSynth91 benchmarks can be used for binary and multi-valued data structures. 

\begin{figure}[!ht]
\begin{center}
			\includegraphics[width=0.4\textwidth]{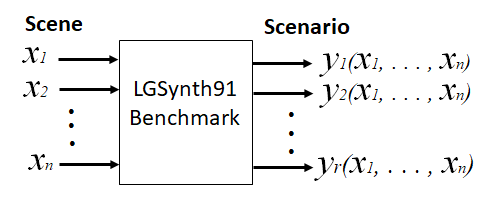}
\caption{Adaption of the LGSynth91 benchmark to the stress propagation in small team: assignments of inputs  are interpreted as scenes, and  outputs  are interpreted as scenarios.}\label{fig:benchmark}
\end{center}
\end{figure}

Adaptation of the LGSynth91 benchmarks for purposes of stress propagation problem includes 
\begin{enumerate}
	\item Problem formulation in terms of computational logic,
		\item Interpretation of any assessment of input variables $(x_i=\sigma_i, \sigma_i\in\{0,1\},i=1,2, \ldots , n$ as scene description of $n$ teammates (Fig. \ref{fig:benchmark}),
			\item  Interpretation of outputs $y_j(x_i), j=1,2, \ldots , r)$ as $r$ scenarios of stress propagation (Fig. \ref{fig:benchmark}), and 
				\item Simulation using proposed model and a given benchmark.
\end{enumerate}


Sample of benchmarks is given in Table \ref{tab:benchmark}. This table  contains the following data:

\begin{itemize}
	\item [$-$]the 1st column `Name': the benchmark name;
		\item [$-$]the 2nd column `Teammate': the number of teammates (originally the number of inputs); 
			\item [$-$]the 3rd column `Scenario': the number of available scenarios (originally the number of outputs);
				\item [$-$]the 4th column `Node': the number of nodes in DD;
					\item [$-$]the 5th column `Path': the number of generated paths.
\end{itemize}
 
For example, the benchmark `dc1' can be used for modelling stress propagation in seven scenarios  of four teammates. They are   generated using DD with 22 nodes. Overall $7\times 2^4$ unique paths represent stress propagation.

\section{Discussion}\label{sec:Discussion}

We developed a novel stress propagation model for first-responders and tactical operators. In contrast to the well-known composite models, the proposed model is based on a set of user-centric principles such as technological advancing in all parts of the model, possibilities for publicly available software packages and benchmarks, and transparency of modelling process, e.g., \cite{[Panganiban-2020]}. 
This  composite model combines the advanced techniques from computational logic, computational epidemiology, and computational genomics, and suggests a number of benefits for researchers and practitioners as described below.

\begin{enumerate}
	\item Development of a composite model is needed since the known  models, e.g., \cite{[Pandey-2016],[Scata-2018],[Urizar-2016],[Xiong-2021]}, do not satisfy the requirements of new  realities such as  teaming of humans and autonomous machines.  In addition, wearable technologies based on a personalized collection of physiological and psychological data, e.g., \cite{rodrigues2018wearable},    provide new possibilities for mission monitoring and control.
		\item The proposed model of stress propagation was created using the \textbf{``standard'' technological  components}, such as software tools including the DD package, message-passing package, and test-benches. 
			\item The proposed model was created using the \textbf{advanced and periodically updated} methodological and computational components and software packages.
			\item A beneficial property of the proposed model is its two-level management hierarchy: visual (graph) and operational (algorithmic) (Fig. \ref{fig:Model_Composite}).

\end{enumerate}

These benefits are reflected  in the computational guidelines, in particular:

\begin{itemize}
	\item [$-$] The mechanism of the decision table provides a \textbf{deep profiling} of a given first responder team accordingly to the mission scenarios and particular scenes. Visualization and graphical interpretation are the attractive features of this mechanism.
		\item [$-$] \textbf{Scalability limitations} of the decision table are overcome using the DDs. An arbitrary mission scenario and scene can be generated by such a DD.
			\item [$-$] The probabilistic  reasoning was selected in order to provide a \textbf{reasoning about the mission}. The computation on the factor graph is viewed as the most general form of a probabilistic modelling \cite{[Koller-2009]}.
				\item [$-$] \textbf{Adaptability} is an inherent property  of the composite model (Fig. \ref{fig:Model_Composite}): 1) the states of teammates and combatant union can be multivalued instead of binary \cite{[Yanush_Decision_Diagram_2006]}, 2) the inference and reasoning based on factor graphs can be extended using Bayesian networks, Dempster-Shafer networks, fuzzy networks, Markov networks, and Markov random fields \cite{[Rohmer-2020]}.
\end{itemize}

These recommendations  provide information to  practitioners who  train teammates (humans and autonomous machines), as well as plan and manage  their missions. Information obtained from the simulation is valuable where it is impossible or costly to get the real-world data, e.g.,  
 scenarios that are caused by  a combinations of rare events, and individual behaviour in various potentially possible scenarios and scenes.


The common issue of modelling first responder teaming is that a model has to serve a purpose, there is no general-purpose models. The proposed composite model is a first responder-centric model for stress propagation. There are various soft factors that are out of the scope of this paper; they include potentially irrational behaviour, subjective choices, and complex psychology of human teammates.


\section{Conclusions and future works}\label{sec:conclusions}

The key contribution of this work is the proposed composite model for the stress propagation in first-responder teams.  The proposed model addresses a rich spectrum of  tasks simulation: from two teammates, e.g.,  task execution   depending on  trust  between human and machine teammates,  to dozens teammates of an arbitrary human-machine copy strategy and tactic.   
This is the reason to shift the focus from the simulation of particular cases to an application-centric style as  a general recommendation on how to assess the risks and to model potential scenarios of success or failure of mission. 

We conclude that: 
\begin{enumerate}
	\item The proposed model is computationally affordable and can be implemented for both small-scale and large-scale problems because of the existing software packages;
		\item The model is versatile, as it can be used to model other phenomena and interventions in first-responder teams, e.g., infections, risk, and trust.
\end{enumerate}

Modelling of stressful scenarios is an essential part of the training process. The results of this paper  contribute to the  development of  training tools (e.g., intelligent assistants and games) for first-responders who  operate in extremely stressful  and dangerous environments. 
The primary goal of stress training is to apply the gained skills to the real‐world operational environment. Stress training should be context-specific and designed to provide pre-exposure to the stressful conditions that are likely to be encountered in the operational environment.




\section*{Acknowledgment}
This project was partially supported by the Natural Sciences and Engineering Research Council of Canada through the grant ``Biometric intelligent interfaces'',  and  by the Department of National Defence's Innovation for Defence Excellence and Security (IDEaS) program, Canada.


\end{document}